\documentclass[a4paper,
twoside,10pt]{article}

\input{jgrg21.sty}

\begin{document}
%
\pagestyle{fancy}
\fancyhead{}
  \fancyhead[RO,LE]{\thepage}
  \fancyhead[LO]{K. Kobayashi}                  
  \fancyhead[RE]{Flux vacua in DBI type Einstein Maxwell theory }
\rfoot{}
\cfoot{}
\lfoot{}


\label{P09}    


\title{%
 Flux vacua in Dirac-Born-Infeld type Einstein-Maxwell theory 
}

%

\author{%
  Takuya Maki\footnote{Email address: maki@jwcpe.ac.jp}$^{(a)}$,
  Nahomi Kan\footnote{Email address: kan@yamaguchi-jc.ac.jp }$^{(b)}$,
  \underline{Koichiro Kobayashi}\footnote{Email address: m004wa@yamaguchi-u.ac.jp}$^{(c)}$
  and Kiyoshi Shiraishi\footnote{Email address: shiraish@yamaguchi-u.ac.jp}$^{(c)}$
}
%


\address{%
$^{(a)}$ Japan Women's College of Physical Education, Setagaya, Tokyo 157-8565, Japan \\
$^{(b)}$ Yamaguchi Junior College, Hohu-shi, Yamaguchi 747-1232, Japan\\
$^{(c)}$ Yamaguchi University, Yamaguchi-shi, Yamaguchi 753-8512, Japan
}

\abstract{
We study compactification of extra dimensions in a theory of Dirac-Born-Infeld (DBI) type gravity.
We investigate the solution for Minkowski spacetime with an $S^{2}$ extra space. 
The solution is derived by the effective potential method in the presence of the magnetic flux on the
extra sphere.
We find that, in a certain model, the radius of the extra space has a minimum value independent of the
higher-dimensional Newton constant in weak-field limit.
}

\section{Introduction}
Recently, models including the higher derivative terms are widely studied as a modified version of
Einstein gravity.
Moreover, various works are reported about compactification with an extra space in the higher derivative
gravity (for example \cite{P09_ref1}).

The Dirac-Born-Infeld (DBI) type gravity has been considered by Deser and Gibbons \cite{P09_ref2} and
studied by many authors. It is expected that the nonlinear nature of the model may remove the possible
singularity of spacetime.

In our recent work, we considered a model of Weyl invariant Dirac-Born-Infeld (DBI) type gravity.
This model contains the Weyl gauge field.
It is natural to put the gauge field into DBI type gravity, since originally DBI theory aimed at relaxing
the singularity of the electric field.

In the DBI electromagnetism, the Lagrangian is
\begin{equation}
\mathcal{L} = - \sqrt{- \det \ (\eta_{\mu \nu} + \beta F_{\mu\nu})} .
\end{equation}
In the four-dimensional spacetime they are satisfied that {\small $- \det (\eta_{\mu\nu}+\beta F_{\mu\nu}
) = 1-\beta^2 (E^2-B^2)-\beta^4(E \cdot B )^2$},
\begin{equation}
\nabla \cdot D = \nabla \cdot \frac{E}{\sqrt{1-\beta^2 E^2}}. 
\end{equation}
In the case of the point charge, 
the electric field is $E \propto \frac{1}{\sqrt{r^4 + \beta^2}}$.
The energy density is also finite.

\section{Spacetime metric and flux}
We first consider the theory with the massless gauge field in six dimensions (thus in the flat spacetime,
it seems to be the DBI electromagnetism), and compactification of the extra dimensions.
The Lagrangian of our model is the following:
\begin{equation}
\begin{split}
\mathcal{L} = &-\sqrt{- \det (f^2 g_{MN} - \alpha_1 R_{MN} + \beta F_{MN} )} + (1 -\lambda)f^6 \sqrt{-g},
\end{split}
\end{equation}
where $f$ is a mass scale. 
$\alpha_1$, $\beta$ and $\lambda$ are dimensionless parameters.
$M,N$ range over $0,1,2,3,5$ and $6$.

Note that one can rewrite the Lagrangian as in the form
\begin{equation}
\begin{split}
\mathcal{L} = &-\sqrt{- \det \mathcal{M}_{MN}} + (1-\lambda) f^6 \sqrt{-g} \\ & = - \sqrt{-g} \sqrt{\det
{\mathcal{M}^M}_N} + (1 -\lambda)f^6 \sqrt{-g},
\notag
\end{split}
\end{equation}
where it is satisfied that
\begin{equation}
{\mathcal{M}^M}_N = f^2 {\delta^M}_N - \alpha_1 {R^M}_N + \beta {F^{M}}_N. 
\end{equation}
Now we assume that the spacetime is described by a direct product of four-dimensional spacetime and an
extra space, i.e., the line element is written by
\begin{equation}
ds^2 = g_{\mu\nu}^{(4)} dx^{\mu} dx^{\nu} + g_{mn}^{(2)} dx^m dx^n, 
\end{equation}
where $\mu,\nu=0,1,2,3$ while $m,n=5,6$.
We will omit the index within the parentheses which indicates the dimension, as long as confusion would
not occur.

We suppose that the four-dimensional spacetime is a maximally symmetric space.
The Ricci tensor of the spacetime is expressed as
\begin{equation}
R_{\mu\nu} = \frac{1}{4} R^{(4)} g_{\mu\nu}, 
\end{equation}
where $R^{(4)}$ is the scalar curvature of the four-dimensional spacetime.
For the Minkowski spacetime, $R^{(4)}$ equals to zero.
We adopt $S^2$ as the extra space.
Then we find
\begin{equation}
R_{mn} = \frac{1}{2} R^{(2)} g_{mn} = \frac{1}{b^2} g_{mn}, 
\end{equation}
where $R^{(2)}$ and $b$ are the scalar curvature and the radius of the two-sphere, respectively.

We suppose that the constant magnetic flux penetrates the extra sphere, just as in the model of
Ranjbar-Daemi, Salam and Strathdee (RSS) \cite{P09_ref3}. Namely we set
\begin{equation}
F_{mn}=B \sqrt{g^{(2)}} \varepsilon_{mn},
\end{equation}
where $g^{(2)}= \det g_{mn}$.
The totally antisymmetric symbol $\varepsilon_{mn}$ takes the value $1$ for $(m,n)=(5,6)$.
The strength of flux is rewritten as $B = \tilde{B}/b^2$ where $\tilde{B}$ is a constant determined from
a topological number.
Substituting above ansatze into the Lagrangian, we get the reduced Lagrangian as 
\begin{equation}
\begin{split}
\mathcal{L}_0 =& - \sqrt{- g^{(4)}} (4 \pi b^2) \\
& \times \left\{ \sqrt{\left(f^2 - \frac{\alpha_1}{4} R^{(4)}\right)^4 \left[\left(f^2  - \frac{
\alpha_1}{b^2} \right)^2 + \beta^2 \frac{\tilde{B}^2}{b^4}\right]} - (1-\lambda) f^6 \right\}.
\notag
\end{split}
\end{equation}

In this model, the effective Newton constant $G$ in four dimensional spacetime can be read from the
expansion of the Lagrangian in the small curvature limit,
\begin{equation}
\mathcal{L} = \sqrt{-g^{(4)}} \left[ \text{const.} + \frac{1}{16 \pi G} R^{(4)} + \cdots \right]. 
\end{equation}
Thus we find $(16 \pi G)^{-1}$ for the constant radius $b_0$ as
\begin{equation}
\frac{1}{16 \pi G} = 4 \pi {b_0}^2 f^2 \left( \frac{\alpha_1}{2} \right) \sqrt{(f^2 -
\frac{\alpha_1}{{b_0}^2})^2 + \frac{\beta^2 \tilde{B}^2}{{b_0}^4}}. 
\end{equation}

\section{Four dimensional flat spacetime and compactification}
We seek the solution for the four-dimensional flat spacetime.
According to Wetterich \cite{P09_ref1}, we use the method of the effective potential for a static
solution, instead of solving the equation of motion derived from the Lagrangian directly.
We now define a potential
\begin{equation}
V(y) = y \left\{ \sqrt{(1-\frac{\alpha_1}{y})^2 + \beta^2 \frac{\tilde{B}^2}{y^2}} -(1-\lambda) \right\}
\end{equation}
where $y=f^2 b^2$.
Then the equation of motion and the stability condition are equivalent to 
\begin{equation}
\frac{dV}{dy} \Big|_{y=y_0} = V(y_0) =0,\qquad
\frac{d^2 V}{d y^2} \Big|_{y=y_0} > 0 
\end{equation}
for the solution $y=y_0$.
To make the equations simultaneously satisfied, we must tune the value of $\lambda$ to be specific value
$\lambda_0$.

\begin{figure*}[h]
\centering
\includegraphics[width=7cm, clip]{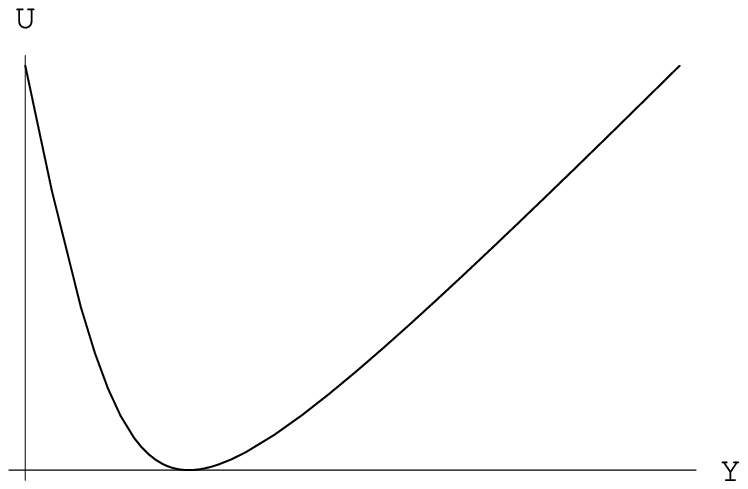}

the DBIE model

\includegraphics[width=7cm, clip]{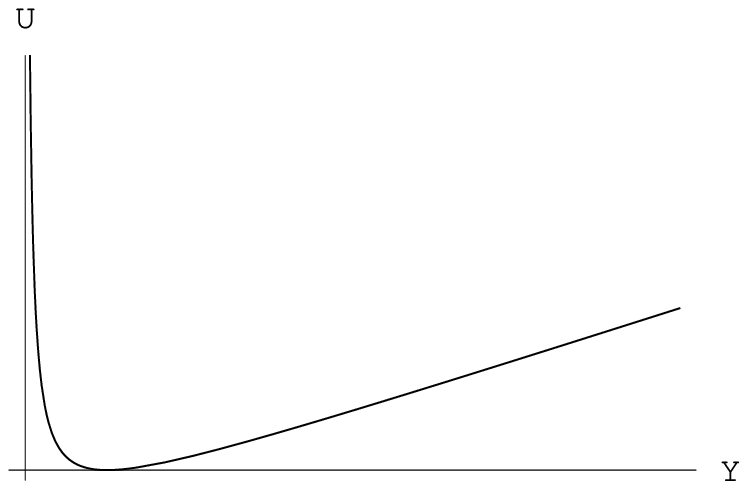}

the RSS model
\caption{Effective potentials in the DBIE model and the RSS model. }
\label{P09_models}
\end{figure*}

We consider further scaling of the variable and parameters become convenient.
This yields a scaled potential
\begin{equation}
\frac{1}{\left| \beta \tilde{B} \right|} V(y) \to U(Y) = Y \left\{ \sqrt{ (1-\frac{A_1}{Y})^2 +
\frac{1}{Y^2} } -(1-\lambda) \right\},
\end{equation}
where $Y=y/|\beta \tilde{B}|$ and $A_1 = \alpha_1/|\beta \tilde{B}|$.
In Figure \ref{P09_models}, we show the $U$-$Y$ graph of potentials in the DBIE model and the RSS model.

Finally we find the solution for
\begin{equation}
\frac{dU}{dY}\Big|_{Y=Y_0} = U(Y_0)=0 \ \  \text{and} \ \ \frac{d^2 U}{d Y^2}\Big|_{Y=Y_0} >0 
\end{equation}
is given by
\begin{equation}
\begin{split}
 &Y_0 = A_1 + \frac{1}{A_1},  \\
 &1-\lambda_0 = \frac{1}{\sqrt{1+{A_1}^2}}
\end{split}
\end{equation}
for $A_1 >0$.
It is interesting to see the minimal value of the radius of $S^2$ is $\frac{\sqrt{|\beta
\tilde{B}|}}{f}$, which is independent of the value of $\alpha_1$.

The inverse of the Newton constant is then given by
\begin{equation}
\frac{1}{16 \pi G} = 2 \pi f^2 |\beta \tilde{B}|^2 \sqrt{1 + {A_1}^2}.
\end{equation}
The squared ratio of the compactification scale and the four-dimensional Planck length is
\begin{equation}
\frac{{b_0}^2}{G} = 32 \pi^2 |\beta \tilde{B}|^3 \sqrt{\frac{(1+{A_1}^2)^3}{{A_1}^2}} \geq 32 \pi^2
|\beta \tilde{B}|^3 \frac{3\sqrt{3}}{2}. 
\end{equation}
Comparing with the result of the RSS model
\begin{equation}
\frac{{b_0}^2}{G}=32 \pi^2 |\beta \tilde{B}|^3 \frac{1}{A_1},
\end{equation}
we find that, in our DBI gravity model, the compactification scale cannot be extremely smaller than the
Planck length, provided that $|\beta \tilde{B}| \sim 1$.

\section{Summary and outlook}
The compactification in the DBI gravity with flux in the extra space has been investigated.
The parameter region which allows the compactification has been revealed.
We have shown that the small couplings attached to the curvature realize similar compactification to that
of the RSS model.

An interesting dependence of the radius of the extra space on the parameter $\alpha_1$ was found in our
model.
This will be of more importance if we consider the parameter as a dynamical variable, or we generalize
our model to include the term such as $\phi^2 R_{MN}$, where $\phi$ is a scalar degree of freedom.

The analysis on stability against perturbation of higher modes, which deforms the spherical shape of the
extra space, is important, even though the analysis on those modes will be complicated because of the
higher-derivative terms in our model.
This issue is left for future works.

Spontaneous compactification to a football-shaped internal space in the presence of a brane is also worth
studying in the framework of the DBI type gravity models, because the higher curvature terms affect the
geometrical aspects of conical or nearly conical points on the compact space.

The cosmological evolution of scale factors in our model is an important subject.
Since the effective potential has a finite value at $b=0$ in our DBI type model, the initial state of the
universe may located at $b=0$.
The simple condition is suitable for quantum cosmology, although the derivative terms in our model make
the canonical approach very complicated.
(The  quantum cosmology of the RSS model was studied by Halliwell \cite{P09_ref4}.)






\begin{thebibliography}{99}
\bibitem{P09_ref1}
C. Wetterich,
Phys. Lett. {\bf B113} (1982) 377.



\bibitem{P09_ref2}
S. Deser and G. W. Gibbons,
Class. Quant. Grav. {\bf 15} (1998) L35 {\tt [arXiv:hep-th/9803049]}.



\bibitem{P09_ref3}
S. Randjbar-Daemi, A. Salam and J. Strathdee,
Nucl. Phys. {\bf B214} (1983) 491.



\bibitem{P09_ref4}
J. J. Halliwell,
Nucl. Phys. {\bf B266} (1986) 228.
\end{thebibliography}
\end{document}